\documentclass[usenatbib,usegraphicx]{mn2e}
\usepackage{times}

\newcommand{\kms}{\ensuremath{\mathrm{km\ s^{-1}}}}

\newcommand{\mgb}{\ensuremath{\mathrm{Mg}\,b}}
\newcommand{\mgfe}{\ensuremath{[\mathrm{MgFe}]}}
\newcommand{\hbeta}{\ensuremath{\mathrm{H}\beta}}
\newcommand{\hdf}{\ensuremath{\mathrm{H}\delta_\mathrm{F}}}

\newcommand{\z}{\ensuremath{\mathrm{[Z/H]}}}

\newcommand{\feh}{\ensuremath{\mathrm{[Fe/H]}}}
\newcommand{\enh}{\ensuremath{\mathrm{[E/Fe]}}}

\begin{document}

 
\title[Hot Stars in Old Galaxies]{Hot Stars in Old Stellar
  Populations: A Continuing Need for Intermediate Ages\thanks{The data
  presented herein were obtained at the W.M.~Keck Observatory, which
  is operated as a scientific partnership among the California
  Institute of Technology, the University of California and the
  National Aeronautics and Space Administration. The Observatory was
  made possible by the generous financial support of the W.M.~Keck
  Foundation.}}
 
\author[S.~C. Trager et al.]{S. C. Trager$^1$\thanks{email:
  sctrager@astro.rug.nl, gworthey@wsu.edu, faber@ucolick.org,
  dressler@ociw.edu}, Guy Worthey$^2$\footnotemark[2],
  S. M. Faber$^3$\footnotemark[2] and Alan Dressler$^4$\footnotemark[2]\\
  $^1$Kapteyn Astronomical Institute, University of Groningen, Postbus
  800, NL-9700 AV Groningen, The Netherlands\\ 
  $^2$Department of Physics and Astronomy, Washington State
  University, Pullman, WA 99164-2814, USA\\
  $^3$UCO/Lick Observatory and Department of Astronomy and
  Astrophysics, University of California, Santa Cruz, Santa Cruz, CA
  95064, USA\\ 
  $^4$The Observatories of the Carnegie Institution of Washington, 813
  Santa Barbara Street, Pasadena, CA 91101, USA}

\maketitle

\begin{abstract}
We investigate the effect of a small contamination of hot, old,
metal-poor starlight on the inferred stellar populations of early-type
galaxies in the core of the Coma Cluster.  We find that the required
correction to the Balmer and metal absorption-line strengths for old,
metal-poor stars does not significantly affect the inferred age of the
stellar population when the \hbeta\ strength is large.
Intermediate-aged populations are therefore still needed to explain
enhanced Balmer-line strengths in early-type galaxies.  This gives us
increased confidence in our age estimates for these objects.  For
galaxies with weak Balmer-line strengths corresponding to very old
populations ($t>10$ Gyr), however, a correction for hot stars may
indeed alter the inferred age, as previously suggested.  Finally, the
inferred metallicity \z\ will always be higher after any correction
for old, metal-poor starlight than without, but the enhancement ratios
\enh\ will strengthen only slightly.
\end{abstract}

\begin{keywords}
galaxies: stellar content --- galaxies: ellipticals and
lenticulars --- galaxies: clusters: individual (Coma)
\end{keywords}

\section{Introduction}
\label{sec:introduction}

Stellar population analysis offers a powerful, if difficult to
interpret, method of understanding the formation histories of nearby
early-type galaxies \citep*[see][for just a few
examples]{Rose85,G93,T00b,CRC03,Mehlert03}.  This analysis relies
primarily on the comparison of a hydrogen Balmer absorption-line
strength to a metal absorption-line strength (or a combination of
metal lines) to break the age--metallicity degeneracy \citep{W94}, as
the Balmer lines are (non-linearly) sensitive to the temperature of
the main-sequence turnoff and the metal lines are sensitive to the
temperature of the red giant branch.  One can therefore determine
accurate ages for the old stellar populations found in early-type
galaxies.  However, other hot star populations such as blue
horizontal branch stars or blue straggler stars can significantly
increase the observed Balmer-line strengths of old stellar populations
\citep*[see,
e.g.,][]{BFGK84,Rose85,RT86,Rose94,dFPB95,MT00,LYL00,T00a}.

In this paper we explore the effect of a specific kind of hot star
population, that is, old, metal-poor populations containing blue
horizontal branch stars, on the inferred stellar population ages and
compositions of early-type galaxies in the Coma Cluster.  These
galaxies appear have significant intermediate-aged populations due to
their enhanced Balmer lines.  We use blue indexes first described by
\citet{Rose85,Rose94} to determine the level of contamination of the
galaxy spectra by blue horizontal branch (BHB) stars.  We then
subtract model spectra representing populations containing these stars
from the observed spectra and determine ages, metallicities, and
enhancement ratios from the residual spectra.  Finally these stellar
population parameters are compared with those determined from the
observed spectra to quantify the effect of a contaminating population
of old, metal-poor stars on the spectra of early-type galaxies.

Throughout this paper, we refer to populations with metallicities
$\z\leq-1.5$ as `metal-poor' (and thus possessing BHB stars) and
populations with ages $1\la t\la10$ Gyr as `intermediate aged'.

\section{Data}
\label{sec:data}

The line strengths discussed in this paper are derived from multi-slit
spectra of twelve early-type galaxies in the Coma Cluster, centred on
the cD galaxy NGC 4874, taken with the Low-Resolution Imaging
Spectrograph \citep[LRIS:][]{LRIS} on the Keck II 10-m Telescope.
Details relevant to the current study are summarised here; for a
complete description of the acquisition, reduction, and calibration of
these spectra and the extraction of Lick/IDS absorption-line strengths
we refer interested readers to Trager, Faber \& Dressler (in
preparation; hereafter TFD05).

\subsection{Observations}
\label{sec:observations}

Spectra were obtained in three consecutive 30-minute exposures on 7
April 1997 UT with the red side of LRIS, with seeing
$\mathrm{FWHM}\approx0.8$ arcsec, through clouds.  A slit width of 1
arcsec was used in conjunction with the 600 line $\mathrm{mm^{-1}}$
grating blazed at 5000 \AA, giving a resolution of 4.4 \AA\ FWHM
($\sigma=1.9$ \AA) and a wavelength coverage of typically 3500--6000
\AA, depending on slit placement.  Spectra of Lick/IDS standard G and
K giant stars and F9--G0 dwarfs \citep{WFGB94} were observed on the
same and subsequent nights through the LRIS 1 arcsec long slit using
the same grating to be used for calibration to the Lick/IDS system
(see Sec.~\ref{sec:lick} below).

Individual two-dimensional spectra of each galaxy were extracted from
the multi-slit images after standard calibrations (overscan
correction, bias removal, dark correction, and flat field correction),
mapping of the geometric distortions, wavelength calibration, and sky
subtraction\footnote{For NGC 4874, which filled its slitlet, and for
D128 and NGC 4872, whose spectra were contaminated by that of NGC
4874, sky subtraction was performed first using the `sky' information
at the edge of their slitlets and then corrected by comparing this sky
spectrum to the average sky from all other slitlets.  The excesses in
these slits were added back into the final extracted spectra.} were
performed following the methodology of \citet{Kelson03}.  Both
individual one-dimensional spectra and variance-weighted, combined
spectra were then extracted from the two-dimensional spectra.  In
order to simulate an equivalent circular aperture to match with other
line strength work in the Coma Cluster, the extracted spectra were
weighted by distance from the object centre.  For the present study
the spectra were extracted with an equivalent circular diameter
aperture of 2.7 arcsec, matching the Lick/IDS galaxy aperture
\citep{TWFBG98} and the fibre diameter of the large sample of line
strengths of Coma galaxies of \citet{M02}.  Finally, the spectra were
flux-calibrated using observations of spectrophotometric standard
stars.

\subsection{Line strengths on the Rose system}
\label{sec:rose}

\begin{table}
  \caption{Observed line strengths of early-type galaxies in the Coma
  Cluster}
  \label{tbl:linestrengths}
  \begin{tabular}{lccccccr}
    \hline
     & Ca\,{\sc ii} & Hn/Fe & \hbeta & \mgb & Fe5270 & Fe5335 & 
     \multicolumn{1}{c}{[O\,{\sc iii}]} \\
    Name & $\sigma$ & $\sigma$ & $\sigma$ & $\sigma$ & $\sigma$ &
     $\sigma$ & \multicolumn{1}{c}{$\sigma$} \\
    \hline
D127    &1.165&1.016&1.878&3.963&3.094&2.761&0.096\\
        &0.014&0.005&0.067&0.072&0.085&0.088&0.048\\
D128    &1.179&1.001&1.855&3.564&2.788&2.542&0.028\\
        &0.012&0.003&0.043&0.044&0.054&0.058&0.030\\
D154    &1.178&0.989&2.147&3.359&2.723&2.701&0.013\\
        &0.034&0.012&0.122&0.124&0.151&0.157&0.085\\
D157    &1.173&1.012&1.817&3.940&2.839&2.670&0.054\\
        &0.009&0.007&0.052&0.052&0.063&0.065&0.036\\
D158    &1.154&0.959&2.237&3.114&2.484&2.309&$-$0.007\\
        &0.021&0.009&0.088&0.089&0.107&0.115&0.061\\
GMP 3565&1.178&0.953&2.297&2.850&2.363&2.287&0.087\\
        &0.074&0.016&0.198&0.205&0.247&0.264&0.140\\
NGC 4864&1.197&1.028&1.738&4.716&2.907&2.889&0.074\\
        &0.012&0.005&0.039&0.041&0.048&0.049&0.027\\
NGC 4867&1.092&0.992&1.938&4.589&2.909&2.897&0.146\\
        &0.017&0.006&0.031&0.032&0.038&0.039&0.022\\
NGC 4871&1.160&1.020&1.832&4.442&2.957&2.921&0.120\\
        &0.009&0.003&0.034&0.036&0.044&0.045&0.024\\
NGC 4872&1.165&1.023&1.781&4.598&2.923&2.836&0.104\\
        &0.018&0.009&0.025&0.026&0.031&0.031&0.017\\
NGC 4873&1.117&0.976&1.867&4.457&2.827&2.663&0.094\\
        &0.011&0.007&0.042&0.044&0.053&0.054&0.029\\
NGC 4874&1.192&1.049&1.572&4.954&3.019&3.035&0.110\\
        &0.012&0.005&0.039&0.041&0.050&0.048&0.027\\ 
\hline
  \end{tabular}

\medskip
All indexes observed within an aperture of 2.7-arcsec diameter.
\end{table}

\citet{Rose85,Rose94} has developed an absorption-line strength
system, based on a combination of line-depth ratios and equivalent
widths, which provides powerful tools for decomposing the hot-to-cool
and giant-to-dwarf star ratios in composite populations.  The Rose
indexes of interest here are Ca\,{\sc ii} (= Ca\,{\sc ii}
H+H$\epsilon$/Ca\,{\sc ii} K) and Hn/Fe
($=\langle\mathrm{H}\theta/\lambda3859 + \mathrm{H}\delta/\lambda4045
+ \mathrm{H}\gamma/\lambda4325\rangle$, \citealt{CR98}), which
together indicate the presence of hot stars in old, metal-rich stellar
populations \citep{Rose85,Rose94,CRC03}.  

The LRIS spectra were first smoothed to a common velocity dispersion
of $230\,\kms$ as suggested by \citet{CRC03}, except for NGC 4874,
which was left unsmoothed at its intrinsic velocity dispersion of
$271\,\kms$.  Indexes were then computed by finding the minimum
intensity of each absorption line using a cubic spline interpolation
of a small region around the centre of each line and then dividing the
appropriate combination of lines to determine the index value.  For
example, the $\mathrm{H}\delta/\lambda4045$ index is determined by
finding the minimum intensity of the $\mathrm{H}\delta$ absorption
line and dividing it by the minimum intensity of the Fe\,{\sc I}
$\lambda4045$ absorption line.  Indexes were determined independently
from each of the three exposures after smoothing and the mean and
sample standard deviations were used as the final index value and
error, respectively.  Within the errors of the two Rose indexes of
interest (Hn/Fe and Ca\,{\sc ii}), the indexes determined from the
variance-weighted combined spectra and from the mean of the individual
exposures are identical.  However, the uncertainties determined from
the standard deviations of the indexes determined from the individual
exposures appear to be more reliable than those determined from
estimation of the photon noise in the combined spectra
\citep[cf.][]{CRC03}.  Rose line strengths and errors for the current
sample, measured in a synthesised circular aperture of 2.7-arcsec
diameter, are presented in Table~\ref{tbl:linestrengths}.

\begin{figure*}
\includegraphics[width=178mm]{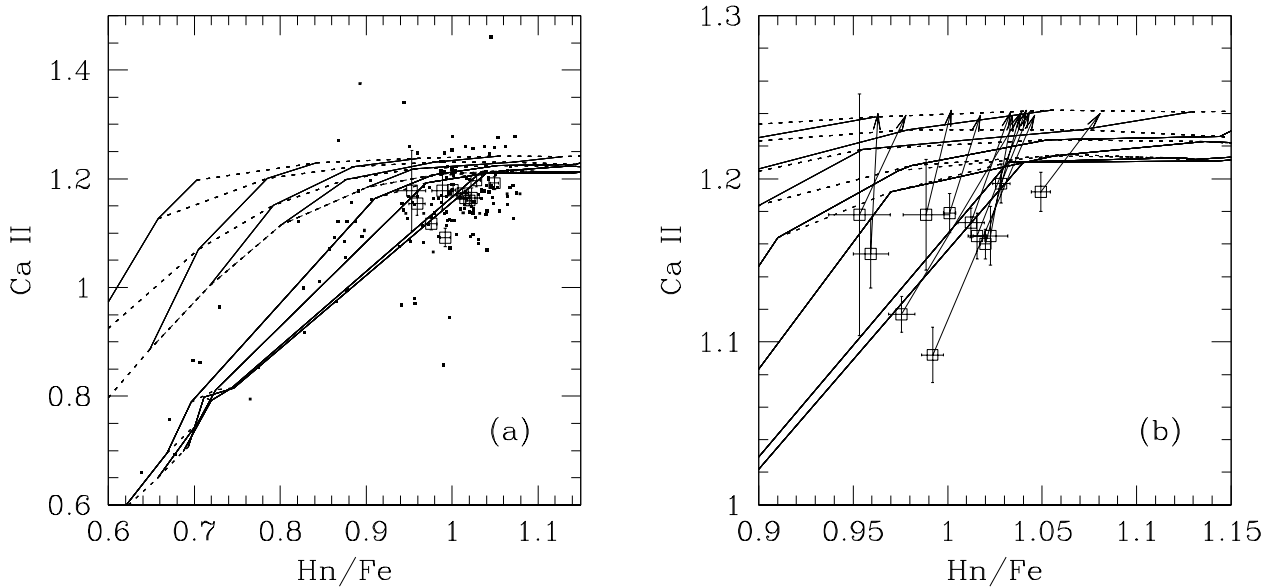}
\caption{The hot-star sensitive Hn/Fe--Ca\,{\sc ii} diagnostic diagram
\citep[cf.][]{CRC03}.  In this figure, grids are taken from the
spectral models of \citet{W94}, where solid lines are isochrones
(constant age) and dotted lines are isofers (constant metallicity).
For old, metal-rich populations Ca\,{\sc ii} saturates at a value of
1.24, while Hn/Fe continues to increase with age.  Galaxies with
Ca\,{\sc ii} below the saturated value are contaminated with hot
stellar populations, most likely metal-poor stars
\citep{Rose85,Rose94}.  In both panels open squares represent Coma
galaxies, observed through a 2.7-arcsec diameter aperture.  (a) The
distribution of field and Virgo Cluster \citep[small filled
points][]{CRC03} galaxies in the Hn/Fe--Ca\,{\sc ii} diagram.  Note
the close overlap between the majority of field, Virgo, and Coma
galaxies in this diagram, confirming that we are on a Rose-like
``system'' (see text).  (b) To demonstrate the amount of contamination
from a hot-star population, we have subtracted off a model spectrum of
a 17 Gyr old, $\z=-1.5$ dex population from our Coma galaxies. To
match the asymptotic value of Ca\,{\sc ii}, a fraction of 4--22 per
cent of the light at 4000 \AA\ is required to be in a metal-poor
population in each galaxy (Table~\ref{tbl:tze}).  Arrows point from
the observed to the corrected line-strength values.}
\label{fig:hnfeca2}
\end{figure*}

No correction for emission fill-in of the Hn/Fe index has been
attempted.  The fluxes of the H$\alpha$ emission in each galaxy
required to make the correction using the recipe of Appendix A of
\citet{CRC03} are unknown, although we expect that the corrections
will be small: the typical correction to Hn/Fe in \citet{CRC03} is
0.012, which is a correction of less than 1 Gyr for an 8 Gyr-old
population.  The precise value of Hn/Fe is however not used in the
analysis that follows.  

Furthermore, we cannot calibrate our index strengths on to a Rose
`system', as the \citet{Jones96} stellar library used by \citet{CRC03}
is smoothed to a velocity dispersion of 103 \kms\ (the intrinsic
resolution of our stellar spectra is roughly 150 \kms), and no
published index strengths exist for the galaxies in the present study.
However, the location of the Coma galaxies in the Hn/Fe--Ca\,{\sc ii}
diagram is coincident with the distribution of field and Virgo
early-type galaxies in \citet{CRC03}, giving confidence that the
moderate velocity smoothing is sufficient to calibrate the Coma
galaxies on to a Rose-like `system' (Fig.~\ref{fig:hnfeca2}a).

\subsection{Line strengths on the Lick/IDS system}
\label{sec:lick}

The Lick/IDS absorption-line strength system has been developed by
Faber, Burstein, and their collaborators
\citep[e.g.][]{BFGK84,WFGB94,TWFBG98} to determine the stellar content
of early-type galaxies \citep*[as seen in the models of,
e.g.,][]{W94,TMB03}.  For the purpose of breaking the age--metallicity
degeneracy inherent in colours and line strengths of old stellar
populations \citep[see, e.g.,][]{O'Connell80}, the \hbeta, \mgb,
Fe5270 and Fe5335 indexes are among the best-understood and
best-calibrated \citep[e.g.,][]{T00a}; we will use these four indexes
to determine stellar population parameters in this study.

In order to measure Lick/IDS line strengths and then to place them on
the Lick/IDS system, a series of spectral manipulations and
calibrations are required.  For the current observations, the complete
series of steps performed is discussed in detail in TFD05; here we
briefly review the process.  

The first step in determining the line strengths of a galaxy is to
measure its systemic velocity and velocity dispersion.  These are
needed to place the index bandpasses on the spectrum and to calibrate
galaxy line strengths on to the Lick/IDS \emph{stellar} system used by
most stellar population models, including \citet{W94}.  Both of these
quantities are measured following the direct-fitting algorithm of
\citet{KIvDF00} over the rest-frame wavelength range 4200--5100 \AA.

Next, the spectrum is smoothed to the Lick/IDS resolution \citep{WO97}
using a variable-width Gaussian filter.  The Lick/IDS index bandpasses
are then placed on the spectrum and indexes measured in either \AA\ or
magnitudes, depending roughly on the width of the central bandpasses
\citep{TWFBG98}.  Errors are determined from the object spectrum and
its associated variance spectrum.  

For galaxies, a correction for fill-in of \hbeta\ absorption by
emission is required \citep{G93}.  This is determined from the
strength of the [O\,{\sc iii}]$\lambda5007$ line as determined from
the residual spectrum after subtracting the best-fitting spectrum from
the models of \citet{V99} from the observed spectrum\footnote{We use
the \citet{V99} models for the emission correction because we achieve
lower $\chi^2$ values in the spectral fitting (i.e., better fits) when
using these models instead of the \citet{W94} models.  This is due to
the denser coverage in age of the \citet{V99} models with respect to
the \citet{W94} models.}.  The correction to \hbeta\ is then
determined using the [O\,{\sc iii}]$\lambda5007$--\hbeta\ correction
suggested by \citet{T00a}: $\Delta\hbeta=0.6\times\,$[O\,{\sc iii}],
where the [O\,{\sc iii}] index is defined by \citet{G93} and is
positive for emission within the central bandpass.  This correction
never exceeds $0.09$ \AA\ for the present galaxies, or about 4\% in
\hbeta\ strength.

Finally, two corrections are required to bring the galaxy indexes on
to the Lick/IDS stellar system: small offsets resulting from the fact
that the Lick/IDS system is not based on flux-calibrated spectra, and
a velocity dispersion correction to account for the velocity
broadening of the galaxies \citep{G93,TWFBG98}.  The former correction
is performed using observations of the Lick/IDS stars taken in the
run; the latter is performed using the polynomial corrections given in
\citet{TWFBG98}.  The final Lick/IDS index strengths of early-type
galaxies in the Coma Cluster for the four lines of interest and for
[O\,{\sc iii}] are given in Table~\ref{tbl:linestrengths}.

\section{Analysis}
\label{sec:analysis}

Our goal is to determine the fractional contribution of hot stars,
\emph{assumed to arise from the blue horizontal branch stars of an
old, metal-poor population,} to the light at 4000 \AA\ of early-type
galaxies in the Coma Cluster and then to correct their age- and
metallicity-sensitive line strengths for this contamination.  We then
compute the stellar population parameters age $t_{\mathrm{SSP}}$,
metallicity $\z_{\mathrm{SSP}}$, and enhancement ratio
$\enh_{\mathrm{SSP}}$ for each galaxy as observed and after correction
for hot-star contamination.  We can therefore determine the effect of
these stars on the inferred stellar populations of early-type
galaxies.  Here SSP refers to the equivalent \emph{single stellar
population}, that is, a population of stars formed at the same time
$t_\mathrm{SSP}$ with the same chemical composition $\z_\mathrm{SSP}$
and $\enh_\mathrm{SSP}$, with the same line strengths as the galaxy.

\subsection{Models}
\label{sec:models}

We use the models of \citet[hereafter W94]{W94} to analyse the stellar
populations of early-type galaxies, extended to cover non-solar
abundance ratios using the \citet{TB95} response functions as
described in \citet{T00a}.  In particularly, we use the `vanilla' W94
models, with extended horizontal branches at low metallicities,
described in \citet{L96}.  While these models provide line strengths
on the Lick/IDS system, they do not normally provide Rose index values
nor spectra \citep[as in, e.g.,][]{V99,BC03}.  \citet[see also
\citealt{CRC03}]{LW00} extended the W94 models to produce spectra
using the empirical stellar spectra library of \citet{Jones96} and
theoretical stellar spectra generated with the SYNTHE and ATLAS
programs (R.~L.~Kurucz 1995, priv.~comm.).  We use these model spectra
in our analysis of the Hn/Fe--Ca\,{\sc ii} plane
(Fig.~\ref{fig:hnfeca2}) and in our correction of the observed
early-type galaxy spectra for contamination by hot-star light.

To determine stellar population parameters, we fit observed (or
corrected) line strengths to model line strengths using the method
described in TFD05.  Briefly, we determine stellar population
parameters using a nonlinear least-squares code based on the
Levenberg-Marquardt algorithm.  Stellar population models are
interpolated on the fly to produce model indexes which are compared to
the observed (or corrected) indexes \hbeta, \mgb, Fe5270 and Fe5335.
Uncertainties are determined by taking the dispersion of stellar
population parameters from 500 Monte Carlo trials using the errors of
the observed line strengths, assuming that these errors are normally
distributed.

\subsection{Method}
\label{sec:method}

\begin{figure}
\includegraphics[width=84mm]{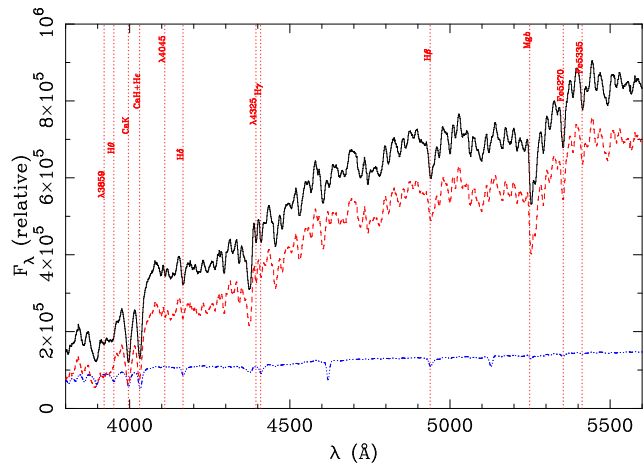}
\caption{The correction of an observed early-type galaxy spectrum for
contamination by an old, metal-poor population containing hot BHB
stars.  The upper (solid) spectrum is the observed spectrum of NGC
4867, an elliptical galaxy in the Coma Cluster.  The lower
(dot-dashed) spectrum is that of NGC 6254 (M10), a globular cluster
with a blue horizontal branch, from the compilation of \citet{SRCM05},
after smoothing, shifting to the systemic velocity of NGC 4867,
normalising at 4000 \AA\ and multiplying by 0.3.  The middle (dotted)
spectrum is NGC 4867 after subtracting the spectrum of NGC 6254.  This
corrected spectrum has a Ca\,{\sc ii} strength of 1.24, as desired.
Absorption lines used in this study are indicated.}
\label{fig:spectra}
\end{figure}

Our method is based on the suggestion of \citet{Rose85,Rose94} and
\citet{CRC03} to use the Hn/Fe--Ca\,{\sc ii} plane to determine the
presence and amount of hot-star light at 4000 \AA\ in early-type
galaxies.  The W94 models predict an asymptotic Ca\,{\sc ii} strength
of 1.24 for \emph{metal-rich} populations with ages
$t_\mathrm{SSP}\ga2$ Gyr (Fig.~\ref{fig:hnfeca2})\footnote{The
\citet{V99} models have a slightly stronger asymptotic Ca\,{\sc ii}
strength of 1.25, which implies slightly higher hot-star fractions by
typically about 1\% of the light at 4000 \AA.  However, we do not use
the \citet{V99} models for the analysis of the impact of the hot-star
population on the stellar population parameters because these models
(1) appear not to have the same flux calibration in the ``blue'' and
``red'' models and (2) have not been modified to account for
enhancement-ratio variations.}. A hot-star spectrum is first smoothed
to instrumental resolution of LRIS and to the velocity dispersion of
the galaxy in question and then subtracted from the observed spectrum
in increments of $f_{\mathrm{hot}}=0.005$ (where $f_{\mathrm{hot}}$ is
the fraction of light coming from the old, metal-poor population at
4000 \AA) until the measured Ca\,{\sc ii} strength in the residual
spectrum reaches the asymptotic old, metal-rich value.  A
demonstration of this process is given in Figure~\ref{fig:spectra} for
NGC 4867.

\begin{table}
  \caption{Line strengths of hot-star populations}
  \label{tbl:models}
  \begin{tabular}{clcccc}
\hline
&&\hbeta&\mgb&Fe5270&Fe5335\\
M&Population&(\AA)&(\AA)&(\AA)&(\AA)\\
\hline
1&17 Gyr, $\z=-1.5$&2.493&1.188&1.029&0.885\\
2&12 Gyr, $\z=-1.5$&2.814&1.185&0.896&0.778\\
3&NGC 6254 (M10)   &2.200&0.687&0.902&0.877\\
\hline
  \end{tabular}

\medskip
M is the model number in Table~\ref{tbl:tze}.
\end{table}

We choose three baseline hot-star models, whose line strengths are
given in Table~\ref{tbl:models}.  The first is a 17 Gyr-old single
stellar population model with $\z=-1.5$ dex and therefore an extended
horizontal branch running from red to blue; the assumed $\enh=0$ dex,
that is to say, having the same \enh\ as the calibrating stars
\citep{TMB03}.  As described in \citet{T00a}, the W94 model ages are
older by 10--25 per cent than models based on isochrones from the
Padova group \citep{Padova}; this 17 Gyr-old model is equivalent in
its line strengths to a 15 Gyr old model from, say, \citet{TMB03}.
The second is a younger, metal-poor single-stellar population with an
age of 12 Gyr, $\z=-1.5$ dex, and $\enh=0$ dex, also with an extended
horizontal branch.  The third is the observed spectrum of the globular
cluster NGC 6254 (M10) from the compilation of
\citet{SRCM05}\footnote{http://www.astro.virginia.edu/~rps7v/GCs/intro.html},
which has $\feh=-1.51$ and an extended, blue horizontal branch
\citep[see, e.g.,][]{RASP00}.  The indexes given in
Table~\ref{tbl:models} are derived from the spectrum of \citet{SRCM05}
and corrected using the LRIS-based flux corrections.

We note here that \citet{SRCM04} present an alternative approach using
the ratio $\hdf/\hbeta$ as a function of Fe4383.  We have not
attempted this method here, as it is dependent on knowing the response
of \hdf\ to $\alpha$-enhancements, which have only just become
available and are currently in the process of being incorporated into
stellar population models \citep{TMK04,KMT05,LW05}.

\section{Results and Discussion}
\label{sec:results}

\begin{figure*}
\includegraphics[width=178mm]{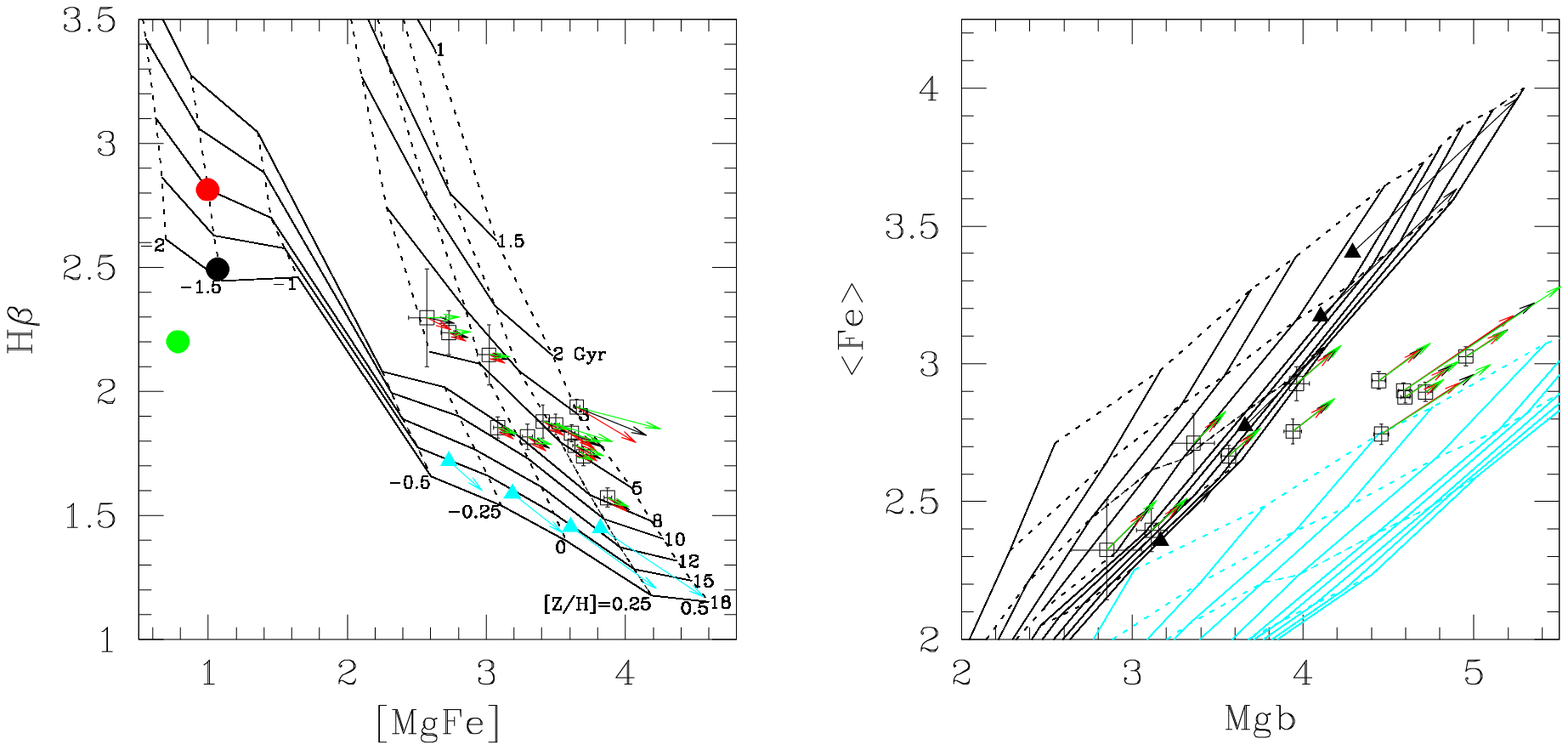}
\caption{The impact of the contamination from metal-poor populations
on the inferred ages and compositions of early-type galaxies (TFD05).
Model grids again come from the W94 models, modified for \enh\ (cf.\
\citealt{T00a}; \citealt{TMB03}; TFD05).  In both panels, solid lines
are isochrones and dotted lines are isofers, as in
Fig.~\ref{fig:hnfeca2}.  In the left panel, the models are for solar
\enh; models with higher \enh\ are nearly identical.  Therefore this
is an appropriate grid from which to visually assess age and
metallicity, although accurate determinations are made in (\hbeta,
\mgb, Fe5270, Fe5335) space.  In the right panel, grids have $\enh=0$
(black) and $+0.3$ (cyan).  Open squares give the observed populations
of the Coma galaxies; arrows point to the population after subtraction
of enough of the old, metal-poor population to bring the Ca\,{\sc ii}
strength to the asymptotic value of old, metal-rich populations
(Fig.~\ref{fig:hnfeca2}).  The black arrows represent the subtraction
of a 17 Gyr old, $\z=-1.5$ dex population (black dot), the red arrows
represent the subtraction of a 12 Gyr old, $\z=-1.5$ dex population
(red dot), and the green arrows represent the subtraction of the
metal-poor globular cluster NGC 6254 (green dot), which has a purely
blue horizontal branch.  The cyan triangles plot the positions of 17
Gyr old populations with metallicities (from left to right) of
$\z=-0.25$, 0, $+0.25$, $+0.5$ dex that have been contaminated by 10
per cent by mass of a 18 Gyr old, $\z=-1.5$ dex population (cyan
arrows) to replicate the experiment of \citet{MT00}.}
\label{fig:hotstars}
\end{figure*}

\begin{table*}
\begin{minipage}{160mm}
  \caption{Stellar population parameters of early-type galaxies in
  the Coma Cluster before and after correction for hot-star
  contamination}
  \label{tbl:tze}
  \begin{tabular}{lrrrcccrrr}
    \hline
    &\multicolumn{3}{c}{observed}&&&&\multicolumn{3}{c}{corrected}\\
    \cline{2-4} \cline{8-10}
    Name & Age (Gyr) & \multicolumn{1}{c}{\z} &
    \multicolumn{1}{c}{\enh} & M & $f_{\mathrm{hot}}$ &
    $f^M_{\mathrm{hot}}$ & Age (Gyr) & \multicolumn{1}{c}{\z} &
    \multicolumn{1}{c}{\enh} \\
    \hline
D127   &$5.3\pm1.4$&$ 0.20\pm0.07$&$0.07\pm0.03$&1&0.09&0.08&$4.8\pm1.0$&$ 0.29\pm0.07$&$0.09\pm0.03$\\
&&&&2&0.08&0.05&$5.1\pm1.2$&$ 0.27\pm0.07$&$0.08\pm0.03$\\
&&&&3&0.11&$\cdots$&$4.5\pm1.0$&$ 0.32\pm0.08$&$0.10\pm0.03$\\
D128   &$8.8\pm1.1$&$-0.08\pm0.03$&$0.03\pm0.02$&1&0.07&0.05&$8.5\pm1.2$&$-0.02\pm0.04$&$0.04\pm0.02$\\
&&&&2&0.06&0.04&$9.0\pm1.2$&$-0.03\pm0.04$&$0.03\pm0.02$\\
&&&&3&0.09&$\cdots$&$8.2\pm1.1$&$ 0.00\pm0.04$&$0.04\pm0.02$\\
D154   &$4.0\pm1.6$&$ 0.08\pm0.12$&$0.01\pm0.05$&1&0.08&0.11&$3.3\pm1.3$&$ 0.19\pm0.12$&$0.04\pm0.06$\\
&&&&2&0.07&0.07&$3.6\pm1.5$&$ 0.16\pm0.12$&$0.03\pm0.06$\\
&&&&3&0.10&$\cdots$&$3.1\pm1.2$&$ 0.22\pm0.12$&$0.04\pm0.05$\\
D157   &$7.6\pm1.2$&$ 0.06\pm0.05$&$0.09\pm0.02$&1&0.09&0.07&$7.2\pm1.5$&$ 0.15\pm0.05$&$0.10\pm0.02$\\
&&&&2&0.08&0.05&$7.6\pm1.4$&$ 0.13\pm0.05$&$0.10\pm0.02$\\
&&&&3&0.11&$\cdots$&$6.8\pm1.4$&$ 0.17\pm0.05$&$0.11\pm0.02$\\
D158   &$4.7\pm0.8$&$-0.11\pm0.06$&$0.05\pm0.04$&1&0.09&0.12&$4.4\pm1.1$&$-0.03\pm0.06$&$0.07\pm0.04$\\
&&&&2&0.08&0.08&$4.6\pm1.1$&$-0.04\pm0.06$&$0.06\pm0.04$\\
&&&&3&0.11&$\cdots$&$4.0\pm1.1$&$-0.01\pm0.07$&$0.07\pm0.04$\\
GMP 3565&$4.5\pm1.7$&$-0.20\pm0.14$&$0.00\pm0.09$&1&0.13&0.20&$4.2\pm2.2$&$-0.08\pm0.18$&$0.03\pm0.10$\\
&&&&2&0.12&0.13&$4.3\pm2.3$&$-0.11\pm0.17$&$0.02\pm0.09$\\
&&&&3&0.18&$\cdots$&$3.7\pm1.8$&$-0.03\pm0.17$&$0.03\pm0.10$\\
NGC 4864&$5.8\pm0.9$&$ 0.30\pm0.04$&$0.21\pm0.02$&1&0.03&0.03&$5.8\pm1.0$&$ 0.33\pm0.04$&$0.22\pm0.02$\\
&&&&2&0.03&0.02&$5.9\pm1.0$&$ 0.32\pm0.04$&$0.21\pm0.02$\\
&&&&3&0.05&$\cdots$&$5.6\pm0.9$&$ 0.34\pm0.05$&$0.22\pm0.02$\\
NGC 4867&$3.0\pm0.2$&$ 0.49\pm0.03$&$0.25\pm0.01$&1&0.22&0.26&$2.4\pm0.3$&$ 0.81\pm0.04$&$0.32\pm0.01$\\
&&&&2&0.19&0.17&$2.9\pm0.2$&$ 0.72\pm0.04$&$0.30\pm0.01$\\
&&&&3&0.30&$\cdots$&$1.9\pm0.2$&$ 0.95\pm0.06$&$0.34\pm0.01$\\
NGC 4871&$4.4\pm0.4$&$ 0.34\pm0.04$&$0.17\pm0.02$&1&0.08&0.08&$4.3\pm0.5$&$ 0.45\pm0.05$&$0.20\pm0.02$\\
&&&&2&0.08&0.05&$4.5\pm0.5$&$ 0.41\pm0.05$&$0.19\pm0.02$\\
&&&&3&0.11&$\cdots$&$4.0\pm0.5$&$ 0.48\pm0.06$&$0.21\pm0.02$\\
NGC 4872&$5.2\pm0.5$&$ 0.29\pm0.03$&$0.20\pm0.01$&1&0.09&0.07&$5.1\pm0.6$&$ 0.39\pm0.03$&$0.23\pm0.01$\\
&&&&2&0.08&0.05&$5.5\pm0.6$&$ 0.36\pm0.03$&$0.22\pm0.01$\\
&&&&3&0.12&$\cdots$&$4.7\pm0.4$&$ 0.43\pm0.04$&$0.24\pm0.01$\\
NGC 4873&$4.8\pm0.8$&$ 0.25\pm0.05$&$0.22\pm0.02$&1&0.17&0.15&$4.4\pm0.7$&$ 0.45\pm0.06$&$0.28\pm0.02$\\
&&&&2&0.15&0.10&$5.0\pm0.8$&$ 0.38\pm0.05$&$0.25\pm0.02$\\
&&&&3&0.23&$\cdots$&$3.9\pm0.7$&$ 0.53\pm0.07$&$0.29\pm0.02$\\
NGC 4874&$8.0\pm0.7$&$ 0.33\pm0.03$&$0.19\pm0.01$&1&0.07&0.05&$8.3\pm0.9$&$ 0.40\pm0.03$&$0.21\pm0.01$\\
&&&&2&0.06&0.03&$8.5\pm0.9$&$ 0.39\pm0.03$&$0.21\pm0.01$\\
&&&&3&0.08&$\cdots$&$8.0\pm0.7$&$ 0.41\pm0.03$&$0.22\pm0.01$\\
\hline
  \end{tabular}

  \medskip
All stellar population parameters are those in a 2.7-arcsec diameter
aperture.  M is the hot-star population given in
Table~\ref{tbl:models}.  $f_\mathrm{hot}$ is the fraction of light at
4000 \AA\ contained in the `hot' (i.e., old, metal-poor) stellar
population as determined from Ca\,{\sc ii};
$f_\mathrm{hot}^M=f_\mathrm{hot}(M/L_B)_\mathrm{hot}/(M/L_B)_\mathrm{total}$
is fraction of the total mass contained in the hot stellar population.
Note that $(M/L_B)$ is unknown for NGC 6254 but is likely to be
slightly higher than that of the 17 Gyr, $\z=-1.5$ dex population
(cf.\ Fig.~\ref{fig:hotstars}).
\end{minipage}
\end{table*}

Figure~\ref{fig:hnfeca2}b shows the results of the correction for
hot-star light on the Hn/Fe and Ca\,{\sc ii} strengths for three of
the Coma galaxies.  Table~\ref{tbl:tze} gives the required fractions
to correct the spectra for hot-star light for all twelve galaxies:
these range from 3 per cent of the total light within the 2.7 arcsec
aperture at 4000 \AA\ for NGC 4864 to 22 per cent for NGC 4867 when
the hot-star light comes from the 17 Gyr old model, with a mean value
of 7.9 per cent and an rms scatter of 2.9 per cent \citep*[here we use
the bi-weight mean and scatter described in][]{BFG90}.  When the 12 Gyr
model is used, these fractions range from 3--19 per cent with a mean
of 7.0 per cent and an rms scatter of 2.6 per cent, and when NGC 6254
is used, the range is 5--30 per cent with a mean of 10.2 per cent and
an rms scatter of 3.7 per cent.

The inferred masses of the metal-poor components are also listed in
Table~\ref{tbl:tze}; the bi-weight mean mass fractions of the
metal-poor component are $8.3\pm6.1$ per cent for the 17 Gyr old
population and $5.4\pm3.9$ per cent for the 12 Gyr old population.
Note that the mass-to-light ratio of NGC 6254 is unknown and therefore
the hot-star mass fractions are also unknown for this case.  These
average mass fractions are consistent with the conclusion of
\cite{WDJ96} that most or all elliptical galaxies have a metal-poor
fraction $\leq 5$ per cent, with a few exceptions depending on the age
of the metal-poor population; GMP 3565, NGC 4867, and NGC 4873 are the
most extreme cases.  We have not fully explored alternative
explanations in this short paper, but GMP 3565 is the most metal-poor
of the galaxies, and thus may have a stronger metal-poor tail if it
has the same abundance distribution as other galaxies shifted to lower
mean abundance. The other two galaxies are candidates to violate the 5
per cent rule, but explanations such as UV-upturn populations,
multiple-age populations, and blue stragglers have yet to be explored.

\begin{table}
  \caption{Stellar population differences after removal of hot-star
  populations} 
  \label{tbl:changes}
  \begin{tabular}{rcccccc}
    \hline
    &$\langle\Delta t\rangle$&$\sigma_{\Delta
    t}$&$\langle\Delta\z\rangle$&$\sigma_{\Delta\z}$&
    $\langle\Delta\enh\rangle$&$\sigma_{\Delta\enh}$\\
    M&frac&frac&dex&dex&dex&dex\\
    \hline
    1&0.942&0.065&0.089&0.033&0.022&0.007\\
    2&1.003&0.046&0.066&0.024&0.014&0.006\\
    3&0.873&0.099&0.115&0.005&0.026&0.012\\
    \hline
  \end{tabular}

\medskip
M is the hot-star population given in Table~\ref{tbl:models}.
$\langle\Delta t\rangle$ and $\sigma_{\Delta t}$ are the mean and rms
scatter in fractional change in the age.
\end{table}

After the appropriate fraction of hot-star light has been subtracted
from the observed spectrum, Lick/IDS line strengths are recomputed
(including velocity dispersion, emission, and systemic corrections).
The results are shown in Figure~\ref{fig:hotstars} as vectors pointing
from the observed line strengths to the corrected line strengths.  As
is clear from this figure, the change in the line strengths point
along vectors of \emph{nearly constant age} with an increase in
metallicity $\z_{\mathrm{SSP}}$ and a slight increase in
$\enh_{\mathrm{SSP}}$ being the major effects.  This is borne out by
the inferred stellar population parameters in Table~\ref{tbl:tze} and
the fractional changes in Table~\ref{tbl:changes}: only NGC 4867
changes its age by more than $1\sigma$.  However, NGC 4867 is
extrapolated far off of the W94 grids after subtraction in all three
cases, and so the inferred ages and metallicities are suspect.

We therefore find that \emph{the presence of even moderate amounts of
light from hot stars does not significantly alter the inferred ages of
early-type galaxies in the presence of intermediate-aged populations,}
provided that the hot-star light comes from an old, metal-poor
population containing blue horizontal branch stars.  We however do not
claim that the ages of early-type galaxies are \emph{unaffected} by
hot stars.  In fact, we reproduce in Fig.~\ref{fig:hotstars} the
result of \citet{MT00} that the \emph{oldest} stellar populations
($t>10$ Gyr) look younger by 2--5 Gyr when ancient, metal-poor
populations are superimposed on ancient, metal-rich populations.
Rather, we suggest that when an intermediate-aged population is
present, i.e., when $\hbeta\ga1.5$ \AA, the influence of hot stars on
the inferred age is nearly negligible \citep[see also][]{TMBM05}.  The
difference between the effect of hot-star light on intermediate-aged
and old populations is mostly due to the change in slope of the W94
grids at old ages and high metallicities in the \hbeta--\mgfe\ diagram
and to the high \hbeta\ strengths of the observed galaxies due,
presumably, to intermediate-aged stellar populations.

The increased inferred metallicity but nearly constant age in the
corrected population can be understood in the context of the Appendix
of \citet{T00b}, which demonstrated that stellar populations add as
vectors in the \hbeta--metal-line spaces (such as \hbeta--\mgfe).  In
effect, the `two' populations in the Coma early-type galaxies--the
metal-rich, intermediate-aged population with high \hbeta\ and high
\mgfe, and the metal-poor, old population with high \hbeta\ (from the
hot blue horizontal branch stars) and low \mgfe--add to produce
observed populations with high \hbeta\ and moderate \mgfe\ strengths.
The strengthening of the \hbeta\ line from the hot stars is
compensated by a dilution of the continuum around the metal lines
(Fig.~\ref{fig:spectra}), which weakens their measured equivalent
widths.  These weakened metal lines combined with the stronger \hbeta\
line serve to preserve the inferred age of the galaxy while lowering
the observed metallicity.

\section{Conclusions}

We have examined the effect of hot (horizontal branch) stars on the
inferred stellar population parameters of early-type galaxies using
observations of twelve \hbeta-strong early-type galaxies in the Coma
Cluster and spectra drawn from stellar population models as well as an
observed spectrum of a metal-poor globular cluster with a purely blue
horizontal branch.  If the hot-star light comes from ancient,
metal-poor populations \citep{Rose85,Rose94,LYL00,MT00,CRC03}
typically contributing $\la10$ per cent of the light at 4000 \AA\ (as
detected in the Ca\,{\sc ii} index), the ages of these galaxies are
not significantly affected by the correction for this hot-star light.
For the oldest, most metal-rich galaxies, this correction can be
significant, as shown by \citet{MT00} and Fig.~\ref{fig:hotstars}.
But this correction is insignificant for the intermediate-aged
populations found in these early-type Coma galaxies, and likely also
for the field and group galaxies studied by \citet{T00b}, many of
which have similarly high \hbeta\ line strengths.  We suggest
therefore that the claim that old, metal-poor stars can `explain away'
the strong \hbeta\ lines in these early-type galaxies
\citep[e.g.,][]{MT00} is overstated.

The presence of blue straggler stars in these galaxies is still a
possibility to explain the enhanced Balmer-line strengths of
early-type galaxies.  However, as discussed by \citet{Rose85} and
\citet{T00a}, populations of blue straggler stars are subject to the
same constraints as other hot-star populations, as blue straggler
stars typically have spectral types around mid-A.  This is the same
colour as the BHB stars that dominate the Balmer-line strengths of the
old, metal-poor populations we have considered here.  Using the same
arguments we have already presented, therefore, we suggest that blue
straggler stars are unlikely to affect the inferred ages of
\hbeta-strong galaxies.  We conclude that \emph{intermediate-aged
populations are still required} to explain the strong \hbeta\ lines in
early-type galaxies.

\section*{Acknowledgments}

The authors wish to recognise and acknowledge the very significant
cultural role and reverence that the summit of Mauna Kea has always
had within the indigenous Hawaiian community.  We are most fortunate
to have had the opportunity to conduct observations from this
mountain.

It is a pleasure to thank D.~Kelson, C.~Maraston, J.~Rose, and
D.~Thomas for many stimulating discussions, R.~Schiavon for making his
spectrum of NGC 6254 available in advance of publication, L.~MacArthur
for a careful reading of an early version of the manuscript, and an
anonymous referee for suggestions that improved the clarity of the
presentation.  Support for this work was provided by NASA through
Hubble Fellowship grant HF-01125.01-99A to SCT awarded by the Space
Telescope Science Institute, which is operated by the Association of
Universities for Research in Astronomy, Inc., for NASA under contract
NAS 5-26555; by a Carnegie Starr Fellowship to SCT; by NSF grants
AST-0307487 and AST-0346347 to GW; by NSF grants AST-9529098 and
AST-0071198 to SMF; and by NASA contract NAS5-1661 to the WF/PC-I IDT.


\begin{thebibliography}{}
\bibitem[\protect\citeauthoryear{Beers, Flynn, \& Gebhardt}{Beers et
al.}{1990}]{BFG90} Beers T.~C., Flynn K., Gebhardt K., 1990, AJ, 100,
32
\bibitem[\protect\citeauthoryear{Bertelli et al.}{1994}]{Padova}
Bertelli G., Bressan A., Chiosi C., Fagotto F., Nasi E., 1994,
A\&AS, 106, 275
\bibitem[\protect\citeauthoryear{Bruzual \& Charlot}{2003}]{BC03}
Bruzual G.~\& Charlot S., 2003, MNRAS, 344, 1000
\bibitem[\protect\citeauthoryear{Burstein et al.}{1984}]{BFGK84}
Burstein D., Faber S.~M., Gaskell C.~M., Krumm N., 1984, ApJ,
287, 586
\bibitem[\protect\citeauthoryear{Caldwell \& Rose}{1998}]{CR98}
Caldwell N.~\& Rose J.~A., 1998, AJ, 115, 1423
\bibitem[\protect\citeauthoryear{Caldwell, Rose \& Concannon}{Caldwell
et al.}{2003}]{CRC03} Caldwell N., Rose J.~A., Concannon K.~D., 2003,
AJ, 125, 2891
\bibitem[\protect\citeauthoryear{de Freitas Pacheco \&
Barbuy}{1995}]{dFPB95} de Freitas Pacheco J.~A., Barbuy B., 1995,
A\&A, 302, 718
\bibitem[\protect\citeauthoryear{Gonz{\'a}lez}{1993}]{G93}
Gonz{\'a}lez J.~J., 1993, PhD Thesis, Univ.\ California Santa Cruz
\bibitem[\protect\citeauthoryear{Jones}{1996}]{Jones96} Jones L.~A.,
1996, PhD Thesis, Univ.\ North Carolina Chapel Hill
\bibitem[\protect\citeauthoryear{Kelson}{2003}]{Kelson03} Kelson,
D.~D., 2003, PASP, 115, 688
\bibitem[\protect\citeauthoryear{Kelson et al.}{2000}]{KIvDF00}
Kelson D.~D., Illingworth G.~D., van Dokkum P.~G., Franx M.,
2000, ApJ, 531, 159
\bibitem[\protect\citeauthoryear{Korn, Maraston \& Thomas}{Korn et
al.}{2005}]{KMT05} Korn, A.~J., Maraston, C., Thomas, D., 2005, A\&A,
in press
\bibitem[\protect\citeauthoryear{Lee \& Worthey}{2005}]{LW05} Lee,
H.-C., Worthey, G., 2005, ApJS, in press
\bibitem[\protect\citeauthoryear{Lee, Yoon \& Lee}{Lee et
al.}{2000}]{LYL00} Lee H.-C., Yoon S.-J., Lee Y.-W., 2000, AJ, 120,
998
\bibitem[\protect\citeauthoryear{Leitherer et al.}{1996}]{L96}
Leitherer C., et al., 1996, PASP, 108, 996
\bibitem[\protect\citeauthoryear{Leonardi \& Worthey}{2000}]{LW00}
Leonardi A.~J., Worthey G., 2000, ApJ, 534, 650
\bibitem[\protect\citeauthoryear{Maraston \& Thomas}{2000}]{MT00}
Maraston C., Thomas D., 2000, ApJ, 541, 126
\bibitem[\protect\citeauthoryear{Mehlert et al.}{2003}]{Mehlert03}
Mehlert D., Thomas D., Saglia R.~P., Bender R., Wegner G.,
2003, A\&A, 407, 423
\bibitem[\protect\citeauthoryear{Moore et al.}{2002}]{M02} Moore,
S.~A.~W., Lucey J.~R., Kuntschner H., Colless M., 2002, MNRAS,
336, 382
\bibitem[\protect\citeauthoryear{O'Connell}{1980}]{O'Connell80}
O'Connell R.~W., 1980, ApJ, 236, 430
\bibitem[\protect\citeauthoryear{Oke et al.}{1995}]{LRIS} Oke J. B.,
et al., 1995, PASP, 107, 375
\bibitem[\protect\citeauthoryear{Rose}{1985}]{Rose85} Rose J.~A.,
1985, AJ, 90, 1927
\bibitem[\protect\citeauthoryear{Rose}{1994}]{Rose94} Rose J~.A.,
1994, AJ, 107, 206
\bibitem[\protect\citeauthoryear{Rose \& Tripicco}{1986}]{RT86} Rose,
J.~A., Tripicco M.~J., 1986, AJ, 92, 610
\bibitem[\protect\citeauthoryear{Rosenberg et al.}{2000}]{RASP00}
Rosenberg A., Aparicio A., Saviane I., Piotto G., 2000, A\&AS, 145, 451 
\bibitem[\protect\citeauthoryear{Schiavon et al.}{2004}]{SRCM04}
Schiavon R.~P., Rose J.~A., Courteau S., MacArthur L.~A.  2004,
ApJ, 608, L33
\bibitem[\protect\citeauthoryear{Schiavon et al.}{2005}]{SRCM05}
Schiavon R.~P., Rose J.~A., Courteau S., MacArthur L.~A.  2005,
ApJS, in press
\bibitem[\protect\citeauthoryear{Thomas, Maraston \& Bender}{Thomas et
al.}{2003}]{TMB03} Thomas D., Maraston C., Bender R., 2003, MNRAS,
339, 897
\bibitem[\protect\citeauthoryear{Thomas, Maraston \& Korn}{Thomas et
al.}{2004}]{TMK04} Thomas D., Maraston C., Korn A., 2004, MNRAS, 351,
L19
\bibitem[\protect\citeauthoryear{Thomas et al.}{2005}]{TMBM05} Thomas
D., Maraston C., Bender R., de Oliveira C.~M., 2005, ApJ, 621, 673
\bibitem[\protect\citeauthoryear{Trager et al.}{1998}]{TWFBG98}
Trager S.~C., Worthey G., Faber S.~M., Burstein D.,
Gonz{\'a}lez J.~J., 1998, ApJS, 116, 1
\bibitem[\protect\citeauthoryear{Trager et al.}{2000a}]{T00a} Trager,
S.~C., Faber S.~M., Worthey G., Gonz{\'a}lez J.~J., 2000a AJ,
119, 1645
\bibitem[\protect\citeauthoryear{Trager et al.}{2000b}]{T00b} Trager,
S.~C., Faber S.~M., Worthey G., Gonz{\'a}lez J.~J., 2000b AJ,
120, 165
\bibitem[\protect\citeauthoryear{Tripicco \& Bell}{1995}]{TB95}
Tripicco M. Bell R.~A., 1995, AJ, 110, 3035
\bibitem[\protect\citeauthoryear{Vazdekis}{1999}]{V99} Vazdekis A.,
1999, ApJ, 513, 224
\bibitem[\protect\citeauthoryear{Worthey}{1994}]{W94} Worthey G.,
1994, ApJS, 95, 107
\bibitem[\protect\citeauthoryear{Worthey et al.}{1996}]{WDJ96}
Worthey G., Dorman B., Jones L.~A., 1996, AJ, 112, 948
\bibitem[\protect\citeauthoryear{Worthey \& Ottaviani}{1997}]{WO97}
Worthey G. Ottaviani D.~L., 1997, ApJS, 111, 377
\bibitem[\protect\citeauthoryear{Worthey et al.}{1994}]{WFGB94}
Worthey G., Faber S.~M., Gonz{\'a}lez J.~J., Burstein D., 1994,
ApJS, 94, 687
\end{thebibliography}
\end{document}